%% file: ms.tex
\documentclass[12pt,preprint]{aastex}

\slugcomment{Submitted Aug. 25 2004}

\shorttitle{A Resolved Disk around HD~107146}
\shortauthors{Ardila et al.}

\begin{document}

\title{A Resolved Debris Disk around the G2V star HD~107146}


\author{
D.R. Ardila\altaffilmark{1},
D.A. Golimowski\altaffilmark{1},
J.E. Krist\altaffilmark{2},
M. Clampin\altaffilmark{3},
J.P. Williams\altaffilmark{4},
J.P. Blakeslee\altaffilmark{1},
H.C. Ford\altaffilmark{1},
G.F. Hartig\altaffilmark{2},
G.D. Illingworth\altaffilmark{5}
}

\altaffiltext{1}{Department of Physics and Astronomy, Johns Hopkins University, 3400 North Charles Street, Baltimore, MD 21218.}
\altaffiltext{2}{STScI, 3700 San Martin Drive, Baltimore, MD 21218.}
\altaffiltext{3}{NASA Goddard Space Flight Center, Code 681, Greenbelt, MD 20771.}
\altaffiltext{4}{Institute for Astronomy, 2680 Woodlawn Drive, Honolulu, HI 96822.}
\altaffiltext{5}{UCO/Lick Observatory, University of California, Santa Cruz, CA 95064.}

\begin{abstract}

We present resolved scattered-light images of the debris disk around  HD~107146, a G2 star 28.5 pc from the Sun. This is the first debris disk to be resolved in scattered light around a solar-type star. We observed it with the HST/ACS coronagraph, using a 1.8'' occulting spot and the F606W (broad V) and F814W (broad I) filters. Within 2'' from the star, the image is dominated by PSF subtraction residuals. Outside this limit, the disk looks featureless except for a northeast-southwest brightness asymmetry that we attribute to forward scattering. The disk has scattered-light fractional luminosities of $(L_{Sca}/L_*)_{F606W}=6.8 \pm 0.8 \times 10^{-5}$ and $(L_{Sca}/L_*)_{F814W}=10 \pm 1 \times 10^{-5}$ and it is detected up to $6.5$'' away from the star. To map the surface density of the disk, we deproject it by $25^\circ \pm 5^\circ$, divide by the dust scattering phase ($g_{F606W} = 0.3 \pm 0.1$, $g_{F814W} = 0.2 \pm 0.1$) and correct for the geometric dilution of starlight. Within the errors, the surface density has the same shape in each bandpass, and it appears to be a broad (85 AU) ring with most of the opacity concentrated at 130 AU. The ratio of the relative luminosity in F814W to that in F606W has the constant value of $1.3\pm0.3$, with the error dominated by uncertainties in the value of $g$ in each filter. An examination of far infrared and submillimeter measurements suggests the presence of small grains. The colors and the derived values of $g$ are consistent with the presence of dust particles smaller than the radiation pressure limit. The dust generated by the creation of a small planet or the scattering and circularization of a large one, are possible scenarios that may explain the shape of the surface density profile.
\end{abstract}


\section{Introduction}

The presence of dusty disks around main-sequence stars serves as a marker for the existence of planetesimals. Without collisions among planetesimals, or their evaporation, the dust would not be replenished, and it would have disappeared from the system long ago. Thus, debris disks indicate that the planet-formation process is occurring, or has occurred. In particular, the study of disks around low-mass stars illuminates the planet-formation process in the relatively low-radiation environments analogous to the solar system. Resolved images of these systems help to constrain their physical and geometrical properties. Scattered-light images sample the whole disk regardless  of its temperature and this, coupled with the fact that optical and near-infrared detectors have higher angular resolution than far-infrared and submillimeter ones, allows for a rich understanding of the systems. 

Here we present coronagraphic scattered-light images of the disk around the G2 V star HD~107146. This is the first debris disk resolved in scattered light around a solar-like star. The observations presented here address the issue of the diversity of planet formation histories within stars of the same spectral type, like HD~107146 and our sun. To date, only one other debris disk around a non-A star (the M0 star AU Mic, see \citealp{kri04,liu04,kal04}) has been resolved in scattered light. 

The disk around HD~107146 was marginally resolved by \citet{wil04} at 450 $\mu$m. Submillimeter and mid-infrared measurements indicate that its fractional excess luminosity is $f_d=L_d/L_*=1.2 \times 10^{-3}$, due to a dust mass of 0.1-0.4 M$_\earth$, comparable to that of the $\beta$ Pictoris disk. An inner hole (>31 AU) is suggested by the lack of IRAS 25$\mu$m detection. No gas measurements have been reported. The Hipparcos distance of the host star is 28.5 pc, its luminosity is 1.1 L$_\sun$ and age estimates range from 30 to 250 Myr \citep{wil04}. The star is a candidate periodic V-band photometric variable ($P=7$d, \citealp{koe02}).

\section{Observations \label{observations}}

HD~107146 and the PSF reference star HD~120066 were observed with the ACS/HRC on UT~2004 June 5 and UT~2004 July 20. The observations were conducted as part of the guaranteed observing time awarded to the ACS Investigation Definition Team (Prop. ID 9987 and 10330). The images were taken with the F606W (broad $V$) and F814W (broad $I$) filters. For each band and each target, a short direct exposure was followed by a coronagraphic exposure using the 1.8'' mask. The coronagraphic exposures for the target were 2330 sec long in F606W, and 2150 sec long in F814W. For  HD~120066 they were 1990 sec in F606W, and 2250 sec in F814W. 

Here we present a summary of the reduction procedure: a more detailed description, applied to different targets, can be found in \citet{kri04} and \citet{cla03}. The results of the measurements are in Table \ref{tab_res}. To measure the  magnitudes of the target and the PSF reference star, we integrated the direct image flux within a circular aperture $>$6 arcsecs in radius, which includes the saturated stellar core\footnote{Instrument Science Reports, R.L. Gilliland, 05 Jan 2004, www.stsci.edu/hst/acs/documents/isrs}. The transformation between counts/sec and magnitudes was obtained using the STSDAS synthetic photometry package SYNPHOT, which simulates most HST observing configurations. The coronagraphic image of HD~120066 was normalized to, aligned with, and subtracted from the image of HD~107146. Alignment of the images is accurate to within $\pm0.2$ pixels ($\pm0.005$''). The resulting subtracted image was smoothed using a 3$\times$3 median filter and corrected for geometric distortion. Figure \ref{disk} shows the result. To increase the signal-to-noise ratio of the displayed images, we have performed an additional 5$\times$5 median smoothing and re-binned by a factor of two. 

Subtraction errors, caused by mismatches in the colors of the stars or PSF time-variability, dominate the emission within $\sim2''$ from the star, and contribute light at large distances. With only one reference star, we cannot quantify very precisely the magnitudes of these two error sources. The ACS Instrument Handbook\footnote{ACS Instrument Handbook, www.stsci.edu/hst/acs/documents/handbooks/cycle13/cover.html} indicates that a mismatch of three spectral classes would produce subtraction residuals for a star of this brightness of the order of 0.2 $\mu$Jy/arcsec$^{2}$ (or 25.73 mag/arcsec$^{2}$ in the V-band), 5'' away from the target. The actual errors in these observations are likely to be smaller, because the V-I colors of the two stars are the same within errors: for HD~120066 we measure V=$6.32\pm0.05$, V-I=$0.68\pm0.07$. The instrument handbook also indicates that typical time-dependent PSF variations within an orbit (due to variations in focus) will result in errors of the same order. However, Figure \ref{rad_prof} suggest that this may underestimate the error in our observations, as subtraction residuals at 6.5'' occur at the level of $\sim$1 $\mu$Jy/arcsec$^{2}$. 

To quantify the systematic error in the normalization constant between the target and the PSF reference star, we compare the value of this constant obtained by using four different methods: taking the ratio of the stellar flux in each band, taking the ratio of the flux of each star away from the saturated columns, taking the ratio of the number of saturated pixels per unit time in the direct images for each band, and adjusting the value of the constant to produce the cleanest visual PSF subtraction. For each filter, the four methods yield the same normalization constant within 1\%, which corresponds to photometric errors of 0.1 and 0.2 mag arcsec$^{-2}$ in the brightest and faintest regions of the disk, respectively. Given that this is an estimate of a systematic error, in what follows we propagate it linearly to estimate uncertainties in calculated quantities. For the quantities calculated below, this error dominates over all others, including the $\sim$5\% random photometric error.


\section{Results \label{results}}

The circumstellar disk is clearly seen in the subtracted images. Within the limitations of the observations, it appears elongated along the SE-NW direction, and featureless, except for the fact that the SW side is brighter than the NE side.  The excess color of the disk with respect to the star is $\Delta (V-I)=0.4\pm0.3$ (Table \ref{tab_res}). This would be the intrinsic color of the disk if the scattering phase function were independent of wavelength. We also detect an object 6.8'' southwest of the star. The time baseline between the observations taken in the two filters is too short to detect any difference in the relative positions between the star and the object. Smooth elliptical fits subtracted from the object reveal residual spiral structure. Therefore, we believe this to be a faint background spiral galaxy (V=19.4, V-I=1.2). The galaxy is at a different position angle than the offset with respect to the optical position of the star of the SCUBA 450 $\mu$m or the map's extension \citep{wil04}. This suggests that the sub-millimeter measurements are not contaminated by the presence of the galaxy.

The slightly red color of the disk with respect to the star suggests the presence of small grains, which leads us to interpret the NE-SW brightness asymmetry as being due to preferential forward-scattering in an inclined disk. This interpretation is consistent with the slightly elongated shape of the observed disk. By fitting elliptical isophotes to the disk image, we conclude that the disk minor axis has a position angle of $58^\circ \pm 5^\circ$. The bright SW region is symmetric with respect to this axis. The measurements are consistent with the picture of a circular disk inclined $25^\circ \pm 5^\circ$ from the plane of the sky.

\begin{figure}

\plotone{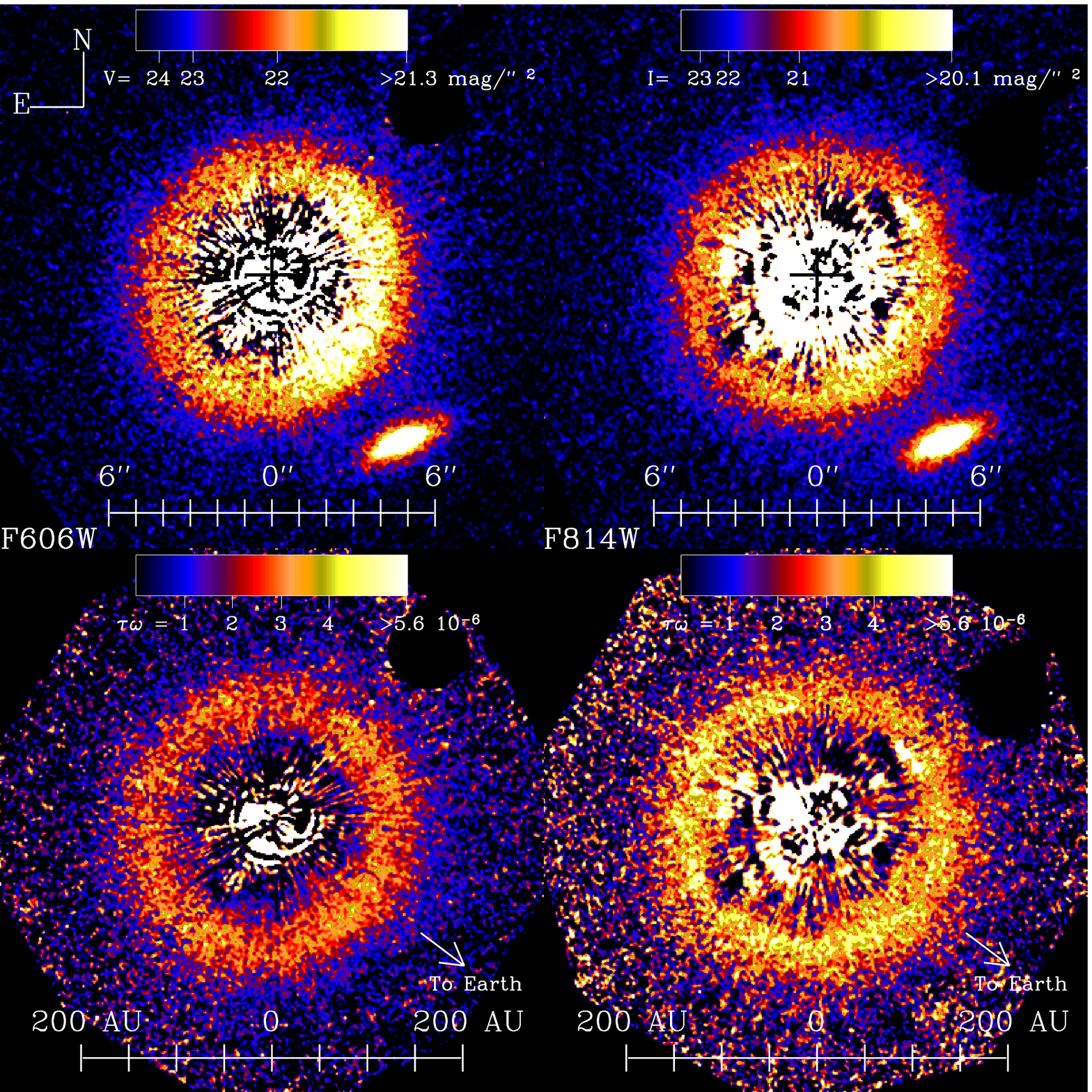}
\caption{\label{disk}The disk around the G2 V star HD~107146. Top left: F606W filter (Broad $V$); Top Right: F814W filter (Broad $I$). To obtain these images, a reference PSF (HD~120066) was normalized, aligned, and subtracted from the coronagraphic images of HD~107146. The 3.0'' spot in the upper right corner has been masked-out. Bottom: Maps of the scattering optical depth $\tau \omega$ (the product of the total optical depth times the albedo), obtained by deprojecting the disk (assuming an inclination of 25$^\circ$ from face on),  multiplying by the distance squared from the stellar position, correcting for forward scattering and dividing by the stellar flux. Left: F606W filter; Right: F814W filter. Note that the bottom row images have the same color scale, but the top row ones do not. The fits files of these images are available in the electronic edition of the journal.}
\end{figure}
\begin{figure}

\plotone{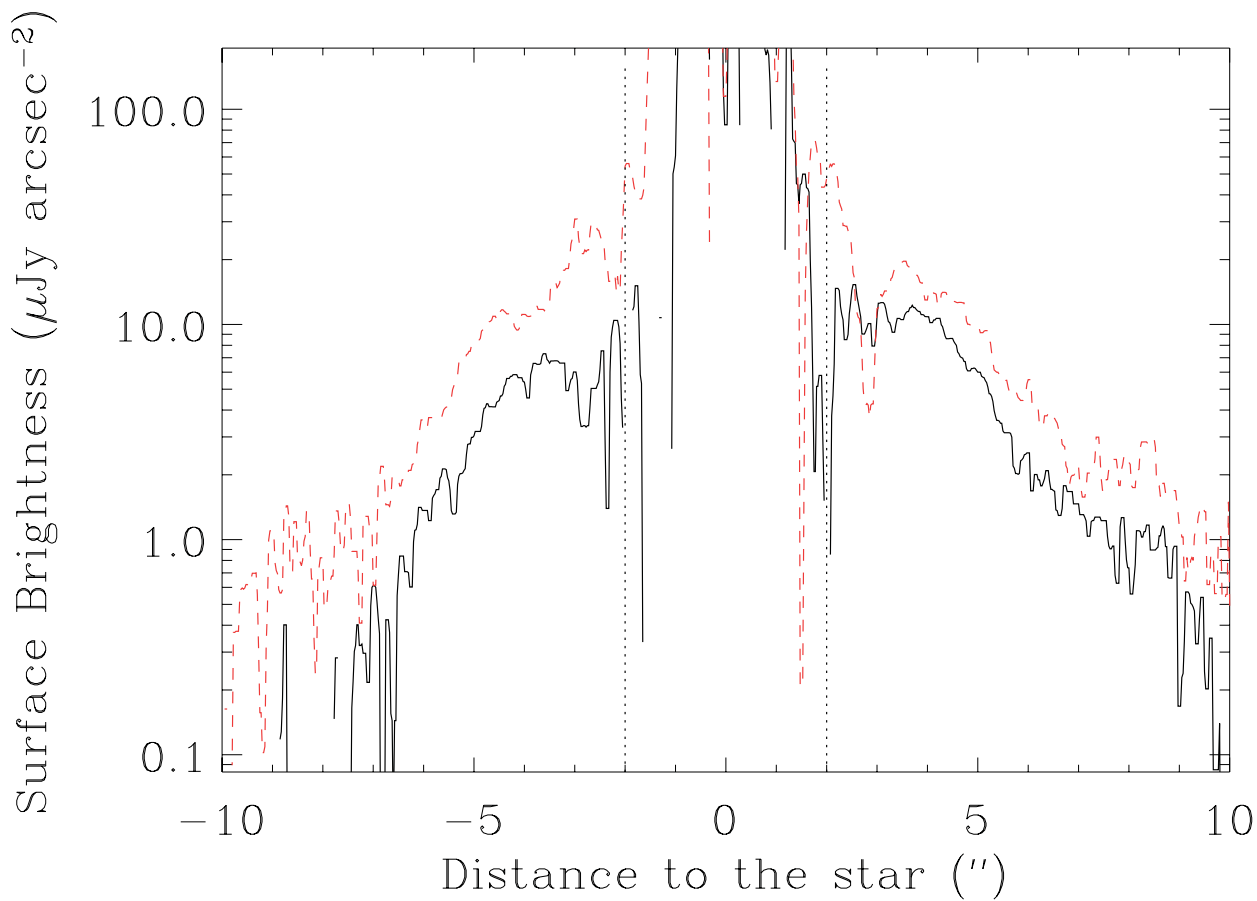}
\caption{\label{rad_prof}Surface brightness as a function of distance, obtained by taking the median of the flux in rectangular boxes 2 arcseconds wide (centered in the stellar position) by 0.125 arcseconds long. The measurements are taken along a path with position angle $58^\circ$ (the disk's minor axis). Positive displacements are towards the SW. The thick black (red dashed) line is the trace in the F606W (F814W) band. The dotted lines are positioned at two arcseconds from the star. This profile suggests that subtraction residuals dominate beyond 6.5'', at the level of $\sim$1 $\mu$Jy/arcsec$^{2}$. 10 $\mu$Jy correspond to V=21.48 or I=20.95.}
\end{figure}

Assuming that the disk is optically thin (as implied by its low $f_d$ value), we can map the surface density \citep{cla03}. We deproject the disk, assuming that it is intrinsically circular, and multiply the resulting image by $r^2$, where $r$ is the distance to the star, to correct for the geometric dilution of starlight. Finally, we divide the deprojected image by a Henyey-Greenstein phase function, adjusting $g$, the scattering asymmetry parameter, until the front- and back-scattering regions have the same brightness (Table \ref{tab_res}). After dividing by the stellar luminosity, the result (Figure \ref{disk}, bottom panels) is a map of the scattering optical depth, which we write as $\tau \omega$, the optical depth times the albedo in the band. The scattering optical depth is proportional to the surface density \citep{wei99, cla03}. To eliminate the galaxy, we fitted its isophotes to a series of ellipses of varying centers, ellipticities, and position angles and generated a model to be subtracted from the image.

\begin{figure}
\plotone{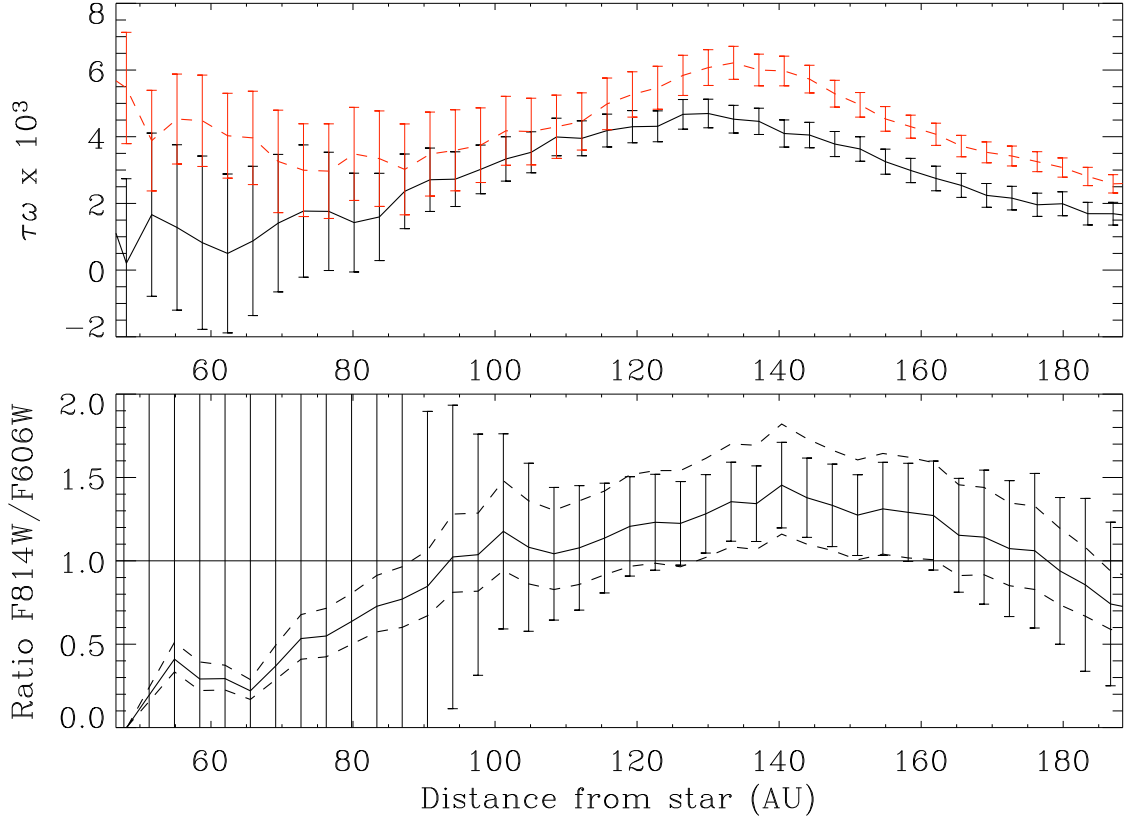}
\caption{\label{colors} Top: Median $\tau \omega$ as a function of distance from the star, obtained in a series of annuli 0.14'' wide, measured from the bottom panels of Figure \ref{disk}. The solid line is the F606W observation ($g_{F606W}=0.3\pm0.1$). The red dashed line is the F814W observation ($g_{F814W}=0.2\pm0.1$). The error bars indicate the range of values obtained if the standard PSF is oversubtracted or undersubtracted by 1\%. Bottom: Ratio of $(\tau \omega)_{F814W}$ to $(\tau \omega)_{F606W}$, as a function of distance from the star. As above, the error bars show the dependence to errors in the normalization. The dashed lines indicate the dependence of the ratio to the value of $g$: the top dotted line is obtained assuming $g_{F606W}=0.2$ and $g_{F814W}=0.3$ and the bottom one is obtained assuming $g_{F606W}=0.4$ and $g_{F814W}=0.1$. These choices produce extremal ratios.}
\end{figure}

Within the level of the errors, the disks shown in the bottom panels of Figure \ref{disk} are azimuthally featureless and they have the same shape. The observed morphology is a broad ring, with maximum opacity at 130 AU and a FWHM of 85 AU. By taking medians of annular sections of the deprojected disks, we map the scattering optical depth as a function of distance (Figure \ref{colors}, top panel). The shape of $\tau \omega$ can be parametrized, as $r^p$, by two power laws: $p=1.6\pm0.5$ (from 80 to 130 AU) and $p=-2.8\pm0.3$ (from 130 to 185 AU). Given the large subtraction residuals within $\sim60$~AU from the star, our observations are not inconsistent with the presence of dust within this radius, nor with a constant optical depth within 80 AU. However, we clearly detect a decrease in the dust opacity within 130 AU: the normalization error would have to be larger than calculated ($>5$\%) for the observations to be consistent with constant opacity within this limit. The difference in the shape of $\tau \omega$ between the two bands is not significant, given the subtraction residuals in the deprojected F814W image.

For the adopted values of $g$, the scattering optical depth is similar in both bands. To measure the color of the deprojected disk, we take the ratio of the two deprojected images and obtain medians of annular sections of the result (Figure \ref{colors}, bottom panel). The color can be reliably determined only between 100 and 180 AU. Between these limits the disk is uniform in color with the mean of the ratio in $\tau \omega$ of the two filters being $1.3\pm0.3$. The uncertainty is determined by the uncertainty in the $g$ values. The uncertainty in the mean is $\pm0.06$. 

\section{Analysis and Discussion\label{anal}}

The outer radius of the disk resolved in scattered light is similar to the submillimeter one. To fit the thermal emission, \citet{wil04} used a single temperature modified blackbody function, in which the emission is $\propto Q_\lambda \ B_\lambda$, with $Q_\lambda=1-exp[-(\lambda_0/\lambda)^\beta]$ and $\lambda_0=100 \mu$m (the characteristic grain size).  They concluded that T=$51\pm4$ K and $\beta=0.69\pm0.15$. 

A detailed model of the thermal infrared emission of the system is beyond the scope of this Letter. Here we note that if we constrain the thermal emission to originate at 130~AU (the radius of maximum brightness in scattered light), and use the values of $\lambda_0$, $\beta$, and $T$ from \citet{wil04}, the disk will not be in thermal equilibrium with the stellar radiation (Backman \& Paresce 1993, Eqn. 1). In the context of this model, the only way to preserve thermal equilibrium is by reducing $\lambda_0$, the characteristic particle size. By keeping $\beta=0.7$ constant, we fit the data with T$\sim45\pm5$~K and $\lambda_0=2\pm1 \mu$m. With $\beta=1$ we obtain T$\sim40\pm5$~K and $\lambda_0=15\pm3 \mu$m. The precise values of $\beta$ and $\lambda_0$ are poorly constrained, as shown by the fact that both sets of parameters fit the thermal data and satisfy thermal equilibrium. For comparison, \citet{den00} found that for $\beta$ Pic, $\epsilon$ Eri, Vega, and Fomalhaut, $\lambda_0=10-100 \mu$m and $\beta=0.8-1.1$. The 25 $\mu$m IRAS non-detection \citep{wil04} and the results of the fit (with $\beta=0.7$) imply that the dust surface density at 130~AU is $\gtrsim$5 times more than at 60 AU, similar to the conclusion reached in scattered light.

Even though it is poorly constrained, the value of $\lambda_0$ suggests the presence of small grains in the disk. The scattering asymmetry parameter, $g=0.2-0.3$, is also consistent with the presence of small grains. Similar values of $g$ are obtained in other debris disks: e.g. $g=0.15-0.25$ for HD~141569A \footnote{Because of a typo, the value of $g$ for the HD~141569A disk is quoted as $g=0.25-0.35$ in \citet{cla03}}  \citep{cla03} and $g=0.4$ for AU Mic \citep{kri04}. For the standard ``astronomical silicate'' \citep{dra84, lao93}, such values indicate the presence of submicron grains, although the actual predicted size is not very sensitive to the value of $g$ \citep{wein01}. The color is also sensitive to grain size. Assuming compact astronomical silicate particles with size distribution $s^{-3.5}$, where $s$ is the particle radius, one can obtain a color ratio between the two bands as large as $\sim1.2$ if the lower limit of the size distribution is $\sim$0.3 $\mu$m. With this dust model a grey disk is observed when then dust particles are $\gtrsim1\mu$m. The ratio of the force due to radiation pressure to the gravitational force is $\beta_{RP}=0.5/s$, with $s$ in $\mu$m (assuming a unity radiation pressure coefficient and a dust density of $\rho_d$=1.25 gr cm$^{-3}$, see \citealp{tak01}). Grains with $\beta_{RP}>0.5$ or $s<1 \mu$m will be expelled from the system in dynamical timescales. In other words, the mean color is consistent with the presence of grains smaller than the radiation pressure limiting size. A more detailed analysis (with a more realistic dust model) is necessary to confirm this result, although a similar situation has been found for the debris disk around HD~141569 (Ardila et al. 2005).


The appearance of a symmetric ring is one of the features of the models by \citet{tak01}. Another is dust segregation: their models predict that the smallest grains present in the system ($\beta_{RP}\sim0.5$) should reside at the largest radii, and larger grains should remain closer to the star. The actual parameters of the segregation depend on the behavior of the gas density at the disk edge. Even in the absence of gas, smaller grains should, in average, reside farther out than larger grains. Over the wavelength span of the two bands, we detect no significant difference in the surface density profile or any systematic color change as a function of distance. Comparison among observations performed over a wider range of wavelengths (for example NICMOS vs. ACS/HRC) are crucial to detect any size separation.

\citet{ken04} show that increased collisions among planetesimals due to the formation of planets with radii larger than 1000 km can generate dust rings. In their models, the outer edge of the ring marks the position at which planets are starting to form.  From their calculations, a planet at 170 AU could grow to the appropriate size if the mass in solids of the disk is between $\sim10$ and $\sim70$ times larger than that of the minimum mass solar nebula. The range is due to the uncertainty in the age of HD~107146. The dust ring maximum fractional luminosity would be $f_d\sim10^{-2}$, larger than the observations. On the other hand, these kind of models produce very axisymmetric structures, similar to those seen in Figure 1. 

An alternative to local planet formation is the migration of a planet formed at a smaller radius. \citet{wei96} show that gravitational interactions among multiple large bodies can scatter one of them into an eccentric large orbit in dynamical timescales. Dynamical friction will induce large eccentricities in smaller bodies, increasing their collision rate and generating broad rings \citep{ken99}. The timescale for this process will depend on the relative masses of the planet and the planetesimals. This scenario would imply that at least two giant planets are present in the HD~107146 system and the presence of the hole reflects the relative paucity of an underlying population of planetesimals, expelled by the (giant) planets. A 50 Myrs old, 10 M$_{Jup}$ planet would have $m_I\sim23 $ mag \citep{bur97}. If the PSF is spread over 4 HRC pixels, each would have a brightness of 16.5 mag/arcsec$^2$. In the F814W subtracted image of HD~107146, the mean surface brightness within 1.8'' is $15.9 \pm 2$ mag/arcsec$^2$, which implies that the planet would not be detected photometrically.
  
Does this disk represent an earlier stage in the evolution of our solar system? The observed dust disk is larger and proportionally much wider than the solar system Kuiper Belt (KB), and it has $\sim$4 orders of magnitude more dust \citep{gre04}. In order to look like the KB, the disk would have to shrink by a factor of $\sim$3 and become narrower by a factor of $\sim$8. This evolutionary path seems contrary to current ideas about the solar system. \citet{gom03} argue that the KB formed closer inward and was pushed out by interactions with Neptune. Additionally, \citet{lev03} argue that the sharp exterior edge of the KB is determined by a 2:1 resonance with Neptune. Defining the edge of the HD~107146 disk at 185~AU implies a planet at 116~AU. There is no dynamical or photometric signature of this object in the scattered light image. We believe therefore that HD~107146 is unlikely to evolve into a system like our own.




\acknowledgments
We wish to thank the team from \citet{wil04} for providing us with the name of their target before publication. ACS was developed under NASA contract NAS 5-32865, and this research has been supported by NASA grant NAG5-7697. We are grateful for an equipment grant from Sun Microsystems, Inc. The Space Telescope Science Institute is operated by AURA, Inc., under NASA contract NAS5-26555.


\clearpage

\clearpage

\input{tab1.tex}

\end{document}

%% file: tab1.tex
\begin{deluxetable}{lcccc} 
\tabletypesize{\small}
\tablecolumns{5} 
\tablewidth{0pc} 
\tablecaption{Measured Quantities\label{tab_res}}
\tablehead{
\colhead{Band} &\colhead{Stellar Bright.} & 
   \colhead{$L_{Sca}/L_*$ \tablenotemark{a}} & \colhead{Disk Mag.\tablenotemark{b}} & 
   \colhead{$g$}\tablenotemark{c} }
\startdata 

F606W& 5.9$\pm$0.3 Jy, V=7.05$\pm$0.05 & $6.8\pm0.8\ 10^{-5}$& V=17.5$\pm$0.1 & 0.3$\pm$1 \\
F814W& 6.9$\pm$0.3 Jy, I=6.35$\pm$0.05 & $10\pm1\ 10^{-5}$& I=16.4$\pm$0.1  & 0.2$\pm$1 \\

\enddata
\tablenotetext{a}{Measured in a circular annulus between 3.5'' and 6.0'' (the brightest part of the disk).}
\tablenotetext{b}{Integrated magnitude in a circular annulus between 3.5'' and 6.0''.}
\tablenotetext{c}{Scattering asymmetry parameter in a Henyey-Greenstein phase function.}                 
\end{deluxetable}